\documentstyle[floats,prl,aps]{revtex}
\begin{document} 
\draft 
\preprint{CWRU-P16-95} 
\title{Does Chaotic Mixing Facilitate $\Omega < 1$ Inflation?} 
\author{ Neil J. Cornish$^1$, David N. Spergel$^{2,3}$ and 
Glenn D. Starkman$^1$} 
\address{$^1$Department of Physics, Case Western Reserve University,
Cleveland, OH 44106-7079} 
\address{$^2$ Princeton University Observatory, Princeton, NJ 08544}
\address{$^3$ Department of Astronomy, University of Maryland,
College Park, MD 20742}
\twocolumn[
\maketitle
\widetext
\begin{abstract} 
Yes, if the universe has compact topology.

\end{abstract} 
]
\narrowtext

%\section{Introduction}

Inflation is currently the most elegant explanation of why the universe
is old, large, nearly flat, homogeneous on large scales and
structured on small scales\cite{inflation}.

One of the most robust predictions of the inflationary scenario,
particularly Linde's chaotic inflationary model, is that $|\Omega-1|$
is exponentially suppressed  and is nearly zero.  This prediction
appears to be contradicted by determinations
of  the Hubble constant, $h \simeq 0.75$,
observations of large-scale structure
that imply that $\Omega_{nr} h \sim 0.25$\cite{peacock}, and stellar ages
that appear to exceed the age of the universe
for these parameters\cite{hogan}.
Here, $\Omega$ and $\Omega_{nr}$ denote the ratio of the 
total energy density to the closure density and
the ratio of the energy density in non-relativistic particles
to the closure density.  This contradiction between theory
and observation has motivated theorists to evoke
a cosmological constant and to explore the possibility
of an open inflationary universe\cite{openinf,openlinde}.

One of the weaknesses of the inflationary paradigm is the
problem of initial conditions for inflation: the pre-inflationary
universe must be somewhat old, somewhat large and
somewhat homogeneous\cite{inflation,homo}.  
%Some authors feel this requirement obviates the successes of inflation, 
%though
In Linde's chaotic inflation model\cite{inflation} and eternal inflation
model\cite{eternal} this is natural, 
but inflation must occur at the Planck scale. 
These initial condition requirements are even more severe in
$\Omega<1$ inflationary models:
if the universe does not inflate enough to appear flat, 
then it does not inflate enough to appear homogeneous\cite{kashlinsky}.

One solution is to have two inflationary epochs:
the first inflation erases all inhomogeneities
and ends with the nucleation of an $\Omega < 1$ bubble, 
which inflates by exactly 69 e-foldings to produce the 
observed universe\cite{openinf}.
%It is a tribute to the various authors' ingenuity 
%that they have devised particle Lagrangians
%in which this double-inflation scenario can be realized
The naturalness of such models has been discussed extensively\cite{openlinde}.
Besides the usual fine tuning of the inflaton potential 
to control the amplitude of density fluctuations, 
and the necessity, in any $\Omega<1$ universe model, 
of arranging that $\Omega$ is relatively close to unity today,
additional tuning is necessary to get both inflations out of one
set of dynamics. 
Open-inflation models have been constructed\cite{openlinde} 
that avoid this latter fine-tuning by using more than one inflaton field.

In this letter, we propose another solution to the
the problem of pre-inflationary homogeneity: 
if the universe is compact, then during the pre-inflationary
period, there is typically sufficient time for 
chaotic mixing to smooth out primordial fluctuations.
Gradients in the energy-density are reduced as $e^{-\kappa d}$, 
where $\kappa$ is the Kolomogorov-Sinai (K-S) entropy of the flow,
and $d$ is the distance the flow travels\cite{chaos}. 
This letter explores this homogenization process, outlines
why small, compact, negatively-curved universes are the most natural,
and concludes with a discussion of the implications of living in 
such a universe.  

%\section{Why Consider Open Universes?}
 
There are several physical and philosophical motivations for
considering compact universes.  
Einstein and Wheeler advocate finite universes
on the basis of Mach's principle\cite{Mach}.
Others argue that an infinite universe
is unaesthetic and wasteful\cite{genetic}, 
because anything that can happen does happen, and an infinite number of times.

Quantum cosmologists have argued\cite{nucleation} 
that small volume universe have small action 
and are therefore more likely to be created.
More intuitively, 
it is more difficult to produce a large universe.
Finally,
a common feature of many quantum theories of gravity
is the compactification of some spacelike dimensions.
This suggests a dimensional democracy, in which
all dimensions are compact,
and geometry distinguishes the large ones from the small ones. 
Positively-curved dimensions remained at or collapsed to
Planck scales in a Planck time, 
while negatively-curved dimensions grew to macroscopic proportions.

The idea of topological or chaotic mixing is also not new\cite{topocosmo}.  
Chaotic mixing has been considered as an alternative to 
inflation\cite{gott,priglock}. However, chaotic mixing does not solve 
the flatness, age,  and monopole problems;
and for it alone to solve the horizon problem, 
the topology scale today would have to be unacceptably small. 
By marrying chaotic mixing to inflation, 
we retain the benefits of inflation and
solve the pre-inflationary initial value problem, 
and in particular the large-scale homogeneity problem for
$\Omega<1$ universes.

Most of the scant attention to non-trivial compact topologies in cosmology 
has focused  on the simplest non-trivial topology of the flat geometry:
a cube with opposite sides identified,
{\it i.e.}~ a 3-torus.
While the universe may be truly flat
($\Omega \equiv 1$, not just $\vert \Omega - 1\vert \ll 1$),
flat manifolds are measure zero in the set of possible 3-manifolds.
Moreover, in flat universes the geometry sets no scale, 
so the size of the fundamental cell of the topology is arbitrary.  
It would be an unnecessary coincidence 
for that scale to be of order the horizon size today.
Positively-curved universes are inherently compact;
however, they typically recollapse on order the Planck time, $10^{-42}s$.
Either inflation must begin at the Planck time, 
as it does in Linde's chaotic inflation models\cite{inflation},
or the universe must be unnaturally flat in order to grow cold enough to
allow inflation to begin.

It has been conjectured that most 3-manifolds are topologically equivalent 
to 3-manifolds of constant negative curvature\cite{thurston}.
This greatly simplifies the description and classification of 3-manifolds. 
The universal covering space of the constant negative curvature geometry
is $H^3$.
The classification of topologies of $H^3$ is then isomorphic to
finding the discrete subgroups of
the group of metric-preserving transitive motions of $H^3$.
This group is isomorphic to the proper Lorentz group $PSL(2,C)$,
which is clear if we think of $H^3$
as a 3-sphere of imaginary radius embedded in 4-dimensional flat space.
Of particular interest are torsion-free subgroups 
as these describe compact 3-manifolds.

While there are an infinite number of compact hyperbolic 3-manifolds, 
they are classifiable in terms of their volumes,
just as 2-manifolds are classifiable by genus.
It has been shown\cite{minvol}
that the volume of any compact hyperbolic 3-manifolds 
is bounded below by $V_{{\rm min}}=0.00082\, R_{curv}^3$, 
where $R_{curv}$ is the radius of curvature.
Many explicit examples have been constructed with small volumes.
Some relatively simple topologies have been constructed 
by identifying the faces of 
the four hyperbolic analogs of the Platonic solids,
the hexahedron, icosahedron, and two dodecahedra\cite{best}.
These typically have volumes in the range $(4-8)\, R_{curv}^3$,
but other examples have volumes as small as
$0.94\, R_{curv}^3$\cite{weeks}.
For our purposes, the volume of the topology is far more important than
the specific form of its identification group.

%\section{Typical Preinflationary Bubble}

We will begin by considering pre-inflationary universes 
that are perturbations around a homogeneous and isotropic solution.   
The Friedman equation,
\begin{equation}
\left({da \over d\eta}\right)^2 = {8\pi G a^4\over 3} 
\left[\left({a_o\over a}\right)^4 \rho_i^{rad}  +
\left({a_o\over a}\right)^2 \rho_i^{curv} + \rho^{vac}\right] 
\end{equation}
governs the evolution of this universe, where
$a$ is the scale factor and $\eta$ is the conformal time.  
We choose $a_0$, the scale factor when the universe is nucleated, 
to be unity without loss of generality.
We have rewritten the usual curvature term in terms of
an energy density, $\rho^{curv}_i = 3 M_{Pl}^2/(8\pi R_{curv}^2)$,
where $R_{curv}$ is the comoving curvature scale.
As we have written explicitly, the vacuum energy density remains constant
as the universe expands, 
the radiation energy density drops as $a^{-4}$,
and the curvature energy density drops as $a^{-2}$. We have neglected
the Casimir energy associated with the finite volume, but it typically
behaves like radiation.

What do we expect for the
properties of a ``typical''
pre-inflationary universe?
Since the only characteristic scale in quantum gravity
is the Planck scale, $M_{Pl}^{-1} =\sqrt {\hbar c/G}$,
we imagine the typical universe
will start with an initial volume, $V_{init} = {\cal C}^3/M_{Pl}^3$,
an initial comoving curvature scale, $R_{curv} = {\cal C}/(\alpha M_{Pl})$,
an initial radiation density, $\rho_i^{rad} = \gamma^2 M_{Pl}^4$, and an
initial vacuum energy, $\rho^{vac} =\lambda M_{Pl}^4/4$.
For compact negatively curved manifolds, $V\equiv \alpha^3 R_{curv}^3$ where
$\alpha > 0.0936$.

We expect ${\cal C}\geq 1$, if only so our
classical treatment of the geometry makes sense.
If nucleation of large universes is suppressed\cite{nucleation}, 
${\cal C}$ should not be too big.

The value of $\gamma$ is also uncertain.  
If $\rho_i^{rad} \mathrel{\raise.3ex\hbox{$<$}\mkern-14mu\lower0.6ex\hbox{$\sim$}} M_{Pl}^4$, then $\gamma^2 \sim 1$;
if the initial energy in radiation is 
$\mathrel{\raise.3ex\hbox{$<$}\mkern-14mu\lower0.6ex\hbox{$\sim$}} M_{Pl}$, then $\gamma^2 {\cal C}^3 \mathrel{\raise.3ex\hbox{$<$}\mkern-14mu\lower0.6ex\hbox{$\sim$}} 1$; 
finally, if the initial energy times the light-crossing time 
is $\sim 1$ ({\it a la} Heisenberg),
then $\gamma^2 {\cal C}^4 \mathrel{\raise.3ex\hbox{$<$}\mkern-14mu\lower0.6ex\hbox{$\sim$}} 1$.
The latter estimate is probably the most appropriate for a finite
volume universe. 
Interpreting the curvature as a source of energy density, 
$\rho_i^{curv} = 3\alpha^2M_{Pl}^4/(8 \pi {\cal C}^2)$, 
we can argue similarly that $\alpha {\cal C} \mathrel{\raise.3ex\hbox{$<$}\mkern-14mu\lower0.6ex\hbox{$\sim$}} \sqrt{8\pi /3}$.

In order for inflation to be consistent with the COBE detections
of large scale temperature fluctuations, 
$\lambda$ must be of order $10^{-15}$, where the
actual value depends on the details of the inflaton potential\cite{inflation}.
In the ``new inflation'' scenario,  
this implies that initially $\rho^{vac}$ 
makes a minor contribution to the total energy density of the universe.  
Thus the universe begins either radiation dominated
if  $\gamma > \sqrt{3/8\pi} (\alpha/{\cal C})$,
or curvature dominated otherwise. During the radiation-dominated period, 
\begin{equation}
\eta = \sqrt{{ 3 \over 8\pi}}\, { ({a}-1)\over \gamma M_{Pl}}  .
\end{equation}
The universe quickly becomes curvature dominated at 
$a_{RC}= \sqrt{8\pi/3}\, \left({\cal C} \gamma/\alpha\right)$.
During the curvature-dominated epoch,
\begin{equation}
\eta-\eta_{RC}=R_{curv}\ln\left[{a \over a_{RC}}\right] .
\end{equation}
Finally, when $\rho^{curv}=\rho^{vac}$
at $a_{CV}=\sqrt{3\alpha^2/(2\pi\lambda{\cal C}^2)}$,
the universe becomes vacuum dominated and 
\begin{equation}
\eta-\eta_{CV}={3 \over  8\pi\lambda M_{Pl}^2}\left({1 \over a_{CV}}
-{1 \over a }\right).
\end{equation}

Consider the evolution of the primordial 
fluctuations during the radiation and curvature dominated phase.
The fluctuations can be expanded as a sum of eigenmodes of the
Laplacian on $H^3$ that satisfy the periodicity conditions of the
identification group.
%However, since the dynamics is chaotic, the time-evolution eigenmodes 
%in the compact topology are not analytic functions, 
%and can only be approximated by an infinite sum of the $H^3$
%eigenmodes\cite{quantchaos}.
If the fluctuations are in a strongly-coupled plasma, 
they will propagate as acoustic waves at a sound speed 
near the relativistic value, $c_s = 1/\sqrt{3}$.
In an open, curvature-dominated universe, these modes
oscillate as free waves that decay only through non-linear
effects\cite{Peebles}.
Classically, the fluctuations in a weakly-interacting field,
such as the inflaton, 
behave as free waves propagating in the expanding background.

All fluctuations will however be suppressed by chaotic mixing.
It is well known to mathematicians\cite{mixing} that geodesic flows 
on a compact negatively-curved manifold (CNCM) are ergodically mixed -- 
bundles of trajectories are stretched and folded like bakers' dough.
A CNCM has an infinite number of unstable eigenmodes.
If initially just one eigenmode is excited, the instability
quickly ensures that all nearby eigenmodes are excited, and the initial
energy spread between them. Since the density of eigenmodes grows
exponentially with wavelength, the system rapidly approaches a state 
indistinguishable from the homogeneous background. That is, the chaotic
mixing acts as an effective (non-collisional) dissipative mechanism.
Given some initial distribution of any propagating field, 
the multi-point correlations decay as $e^{-\kappa d}$, 
where $d$ is the distance a mode of the field has propagated and 
$\kappa$ is the K-S entropy of the flow.  
For a CNCM of volume $V$, $\kappa\simeq V^{-1/3}$.
($L=V^{1/3}$, the ``topology scale,'' 
is approximately the distance across the fundamental cell.)
This description applies exactly to the cosmological situation,
where $\eta$ (or $c_s\eta$) is the comoving distance traveled by a mode
and $\kappa\simeq 1/(\alpha R_{curv}\ln{2})$ is the effective K-S
entropy\cite{priglock}.
%As discussed above, we expect small $\alpha$ to be favored 
%in a quantum creation process for the universe.

One may also be concerned that studying classical flows 
is not appropriate for understanding the behavior of
inherently quantum mechanical fluctuations, 
especially for the weakly-coupled inflaton. This concern is unfounded.
Quantum chaos on CNCMs is known to exhibit the same ergodic mixing seen
in the classical description\cite{quantchaos}. The positive K-S entropy
is manifested as an exponential growth in the number of low energy modes, 
and the instability leads to the exponential decay of the
amplitude of any given mode. In the limit of perfect mixing, 
the wavefunctions become completely random and uncorrelated,
in contrast to the ordered wavefunctions 
found in integrable quantum systems.

The mixing can be understood as follows:
Initially the horizon expands 
to encompass many copies of the fundamental cell (in the covering space).
During this time we expect substantial chaotic mixing.
Once the universe becomes vacuum dominated, however, 
the horizon shrinks rapidly below the topology scale, 
and we expect little or no further mixing.
Post-inflation, the horizon scale grows once again. 
Today it may encompass several fundamental domains, 
leading to a new period of chaotic mixing. 
This mixing would give an additional 
suppression of the large-scale inhomogeneities,
which we neglect below.  Post-recombination mixing has been considered,
and ultimately rejected\cite{postinf}, as a self-contained explanation 
for the isotropy of the cosmic microwave background radiation (CMBR).

From our considerations of the initial state of the universe, we expect
$\rho_i^{rad}/\rho_i^{curv} < 8\pi/(3\alpha^2)$. For $\alpha
\mathrel{\raise.3ex\hbox{$>$}\mkern-14mu\lower0.6ex\hbox{$\sim$}} 1$, almost
the entire pre-inflationary evolution will be curvature-dominated.
Primordial density fluctuations are thus reduced by a factor of
\begin{equation}
{(\delta \rho/\rho)_f\over (\delta \rho/\rho)_i} =
\exp\left[-\kappa c_s\, \Delta \eta \right]
% \exp\left[{-\kappa c_s (\eta_{RC} - \eta_{CV})}\right] 
\simeq \left( {2 \pi \lambda \over 3} {{\cal C}^2\over \alpha^2}\right)^{
                      c_s/(2\alpha\ln{2})} \, .
\end{equation}
For acoustic modes $c_s\simeq 1/\sqrt{3}$.
Fluctuations in the inflaton are more strongly suppressed, since $c_s=1$.
The level of homogeneity necessary for inflation to begin is quite mild;
however, to explain the isotropy of the CMBR on large scales,
evolution during the pre-exponential-inflation phase 
must suppresses temperature fluctuations to at most $10^{-5}$.
Since the total energy in radiation is initially only
${\cal C}^3\gamma^2M_{Pl}$, and since radiation redshifts, 
only the fluctuations in the inflaton need be so severely suppressed.
These $10^{-5}$ temperature fluctuations
translate into inhomogeneities in the inflaton energy density of
$\mathrel{\raise.3ex\hbox{$<$}\mkern-14mu\lower0.6ex\hbox{$\sim$}}
8\times 10^{-5}(1/\Omega - 1)$
for modes of wavelength longer than the curvature scale\cite{openinf}.
Starting from $\delta\rho/\rho\simeq1$ initially
(standard non-linear damping will quickly cut larger fluctuations down
to this size), and saturating the bound
$\alpha {\cal C}\mathrel{\raise.3ex\hbox{$<$}\mkern-14mu
\lower0.6ex\hbox{$\sim$}} \sqrt{8\pi / 3}$,
we find that sufficient mixing can be achieved if
$\alpha \mathrel{\raise.3ex\hbox{$<$}\mkern-14mu\lower0.6ex\hbox{$\sim$}}
2.7$ for $\Omega=0.3$, and $\alpha
\mathrel{\raise.3ex\hbox{$<$}\mkern-14mu\lower0.6ex\hbox{$\sim$}}
3.5$ for $\Omega = 0.1$ for $\lambda = 10^{-15}$.
Thus, small universes, whose ``topology scale'' is comparable to the
curvature scale, will have sufficient time to erase primordial fluctuations.
When the vacuum energy starts to dominate and inflation begins, 
the horizon shrinks below the topology scale and
mixing rapidly terminates.
Therefore, we require that the inflaton is already rolling
%($\dot{\phi} \neq 0$)
by the time the mixing stops,
otherwise quantum fluctuations in the inflaton can lead to large
temperature fluctuations. The inflaton potential must be of a form that
ensures the inflaton comes out of the curvature-dominated epoch rolling at
its full slow-roll velocity\cite{openlinde}.

The exact value of our limit on $\alpha$ relies heavily 
on the value of the K-S entropy, which we have approximated by
its isotropic average.
It relies more weakly on the value of $\lambda$, 
and thus the model used to extract $\lambda$ from the fluctuation spectrum.
For example, for a pure $(\lambda/4)\phi^4$ potential
we would have found $\alpha < 2.4$ for $\Omega = 0.3$, $\alpha < 2.6$ for
$\Omega=0.1$.
Universes with somewhat higher values of $\alpha$ 
should be investigated individually.  

So far, we have restricted ourselves to a constant-curvature background
with perturbations.
This was so we could perform an analytic calculation.
We believe that physically this simplification is unnecessary.
The topology of any 3-manifold is conjectured to be equivalent to
a manifold formed from homogeneous primitives\cite{thurston}.
%Each of these primitives admits only one of eight basic 3-geometries.  
It has been shown that most\cite{thurston} 3-manifolds 
consist of one negatively-curved primitive.
Because compact $H^3$ topologies lead to chaotic mixing, 
we conjecture that most 3-geometries will evolve to
a homogeneous FRW of negative curvature.

%\section{Conclusion}

In summary, if the universe has negative curvature and compact topology, 
the most generic 3-manifold,
then chaotic mixing smooths out primordial inhomogeneities.
We have shown that for compact negatively-curved universes with
volumes on the order of a few Planck volumes at nucleation,
sufficient chaotic mixing will occur to solve the 
large-scale-inhomogeneity/inflationary-initial-value problem.
Thus, unlike in infinite negatively-curved universes, 
two stages of inflation are not required.
We have also argued that this
mechanism will operate in general geometries, causing them to evolve to
a FRW geometry. In the absence of observable curvature, 
this mechanism can solve the initial value problem of inflationary models, 
but is not expected to have observational consequences.
More optimistically, if astronomers have detected the curvature 
of the universe, they may soon detect the effects of its finite size.

As remarked above, most of the attention to compact topologies in cosmology 
has focused on the 3-torus.
The earliest works looked for objects in the sky which could be seen 
in more than one direction, much as you see yourself in a hall
of mirrors.
By using objects that evolve relatively slowly,
lower limits on the topology scale of order $200MpC$ were 
set\cite{objects}.

The effects of compact topology may be more easily seen in the CMBR.
In flat topologies, 
modes of wavelength larger than the identification scale 
(up to $6$ times the topology scale, depending on the topology)
do not exist,
reducing the amplitude of fluctuations on large angular scales.  
This has been searched for in the case of the 3-torus\cite{COBE}.
In negatively-curved compact models there is no such cutoff.
Some authors have proposed searching for periodicity in the pattern
of CMBR hot and cold spots\cite{fagundes}.
More promising, we believe, is to recognize that 
in these small universes the CMBR will
be identified at the intersections of the surface of last scattering (SLS),
as seen by different ``copies'' of the observer (in the covering space).
Here, small means that the diameter of the  SLS
is smaller than the topology scale.
Since the SLS is a sphere, 
these intersections will be circles, 
regardless of the background geometry or topology.
Thus fluctuations in the CMBR would be correlated
on circles of the same radii centered on different points on the sky. 
The existence of these correlated circles will allow us to
search for the existence of topology, 
independent of the particular topology in question\cite{CSS2}.
The COBE/DMR4 data set\cite{bennett}  is currently being analyzed
to search for this signal, but
its signal-to-noise and angular resolution are probably inadequate.
However, data from CMBR satellites planned for the next decade
should allow us to decide definitively 
if there is topology on the scale of the observed universe.

We thank J.R. Bond, L. Krauss, J. Levin, A. Linde and T. Vachaspati
for profitable discussions. 
GDS acknowledges seminal discussions with C. Dyer. 
DNS acknowledges NSF and NASA for support.

\end{document}